# Capable but Not Scalable: Requirements and Interaction Costs of Building Multi-View CAD Datasets for Design Research

Danyal Sarfraz* *first author
Researcher, Department of Human-Centered Design and Technology (HCDT), UNIST

SeEun Lee
Masters Student, Department of Human-Centered Design and Technology (HCDT), UNIST

Ulugbek Vahobjon Ugli Ismatullaev
PhD Candidate, Department of Human-Centered Design and Technology (HCDT), UNIST

JoungHyun Kim
PhD Candidate, Department of Human-Centered Design and Technology (HCDT), UNIST

Abdurakhimov Donierbek
Masters Student, Department of Human-Centered Design and Technology (HCDT), UNIST

Kim, KwanMyung** **Corresponding Author
Professor, Department of Human-Centered Design and Technology (HCDT), UNIST

## 1. Introduction

Image-based AI has become a practical instrument in product design research: models predict aesthetic appeal and generate design candidates from product images (Burnap et al., 2023), recognize 3D shapes from rendered views (Su et al., 2015) and generate CAD models from natural-language prompts (M. S. Khan et al., 2024). Such work depends on datasets, and CAD models are an attractive source, since design teams already produce them.

Turning CAD models into multi-view images is data work, the often undervalued labor on which model quality depends (Sambasivan et al., 2021), and in design research it falls to researchers who need rendered datasets but may not write or adapt rendering scripts. Three parties depend on its output: the design researcher who builds the dataset, the evaluators who rate rendered images as stimuli, and the AI model that learns from them. Recent studies of form indicators and visual quality rely on consistently rendered product images (Ismatullaev & Kim, 2026; Kim & Kim, 2026).

Rendering is not neutral for either consumer. If products are not normalized in size, a model may learn that small parts look small rather than learning their form; if the camera labeled "front" points at a different side of each product, neither a model nor a rater compares like with like; and if appearance correlates with the source of a file rather than its geometry, a model may exploit that correlation, a failure known as shortcut learning (Geirhos et al., 2020), while ratings mix responses to form with responses to lighting and material. We argue that, for design researchers, the problem lies less in what tools can do than in how the work scales: almost every function a controlled dataset needs exists, but it is offered either through code or as a manual step repeated for every product.

**RQ1:** How do current studies convert 3D and CAD models into multi-view images for AI, and what does each approach demand of the person operating it?

**RQ2:** What requirements should a rendering pipeline satisfy so that design researchers can produce datasets that are both AI-ready and meaningful for design research?

**RQ3:** To what extent can design researchers meet these requirements with existing tools without writing code, and which steps remain manual for each product?

The contributions are (1) an analysis of reported rendering practice, (2) fourteen requirements covering both the dataset and the interaction needed to produce it, and (3) an analytic comparison of eleven tools, read through cognitive dimensions, with design implications.

## 2. Background

### 2-1. View-based learning from 3D shapes

Su et al. (2015) showed that a network trained on twelve or eighty rendered views recognizes 3D shapes more accurately than descriptors computed on 3D representations. Product form evaluation adds a different need: corresponding views must have consistent semantic meanings across products, so that front and side views remain comparable.

### 2-2. CAD data and dataset quality

Arjomandi Rad et al. (2024) identify quality assurance and harmonization as recurring barriers to dataset creation in design research. Datasheets for datasets (Gebru et al., 2021) add documentation criteria that apply to the CAD-to-image stage.

### 2-3. Analyzing tools without users

HCI offers analytic methods for judging tools before users are studied. The cognitive dimensions framework (Green & Petre, 1996) names properties that make a tool easier or harder to use. Viscosity is resistance to change; it depends on the operation, and repetition viscosity arises when one change must be repeated in many places (Green & Blackwell, 1998). Hidden dependencies are relationships the tool does not show; visibility and progressive evaluation concern seeing the work and checking it as it develops; closeness of mapping is how directly the tool's terms match the task; and error-proneness is how easily the tool invites slips. Myers et al. (2000) add a tool's threshold, how hard it is to learn, and its ceiling, how much can be done with it, noting that successful tools have tended to be low in both or high in both, and Ko et al. (2004) describe six learning barriers that people face when programming.

## 3. Research Methods

### 3-1. Practice analysis and requirements

We analyzed ten sources that render 3D or CAD models into images for training, evaluating or annotating AI models: two foundational multi-view learning studies (Kanezaki et al., 2018; Su et al., 2015), the widely used Objaverse-XL rendering script (Deitke et al., 2023), and seven CAD-specific studies and datasets from 2024 to 2026 (Doris et al., 2026; Govindarajan et al., 2026; Joglekar et al., 2026; M. S. Khan et al., 2024; M. T. Khan et al., 2025; Pyatov et al., 2026; Shi et al., 2026). Each source was coded on purpose, input format, renderer, number of views, view selection, orientation, normalization and projection from its paper, documentation and public code; attributes a source does not describe were coded as not reported. Dataset requirements combine these findings with principles of machine learning and design research, and interaction requirements follow from the cognitive dimensions; together they are candidate criteria for the reference task rather than an exhaustive account of researchers' needs.

### 3-2. Tool comparison and interaction cost

We rated eleven tools. On the development side, these are BlenderProc2 (Denninger et al., 2023), Kubric (Greff et al., 2022), NVIDIA Omniverse Replicator with its CAD Converter, UnrealCV with Unreal Engine and its Datasmith CAD importer (Qiu et al., 2017), CAD2Render (Moonen et al., 2023), pythonOCC, and two released scripts, the Objaverse-XL renderer (Deitke et al., 2023) and the Ortho2CAD drawing generator (Joglekar et al., 2026). On the design side, they are Blender, KeyShot and Rhino; these and Omniverse were rated from their official documentation. Each tool was rated from public papers, documentation and code against an illustrative reference task for controlled comparison of product form: rendering 300 STEP files of consumer products with a fixed view set, a clay baseline and normalization to the largest dimension, grouped by product with provenance.

The assumed user can configure documented interfaces and run an existing script but does not write or modify code, including visual programs. Each rating describes accessibility for the specified nonprogramming user, not the tool's overall technical capability: A indicates automatic completion for all products after setup; M indicates repeated per-product intervention; and – indicates that no documented route exists without writing or modifying code, or that the function is unsupported. A requirement with several parts is rated by its least supported part; R1, for example, is A only where native STEP import and tessellation that scales with the size of the geometry are both documented, or where, as in Ortho2CAD, B-rep edges are projected without surface meshing. Where a tool cannot import STEP (R1), the remaining requirements are rated assuming the files have been converted to a supported format. The M count indicates potential repetition points, not measured interaction time, action counts, errors or workload. We then traced five actions of the reference task from a documented workflow property to a dimension and a design implication (Section 6-2).

## 4. Current Rendering Practice

Five findings emerge from the existing work.

### 4-1. Rendering is fragmented

Reusable code exists, but it encodes task-specific protocols: the Objaverse-XL script is openly distributed, and Ortho2CAD, CADBench and CADFS release their rendering code (Doris et al., 2026; Joglekar et al., 2026; Pyatov et al., 2026). Although five of the seven CAD-specific sources start from STEP files, their renderers range from Blender (Govindarajan et al., 2026) and pythonOCC (Joglekar et al., 2026) to PyVista and Pyrender (Doris et al., 2026), and several name their renderer only in released code or not at all. The inspected rendering implementations are all code-based, so adapting one requires programming and re-assembling geometric preparation, view assignment, appearance control and output.

### 4-2. View protocols vary widely

The number of views per object ranges from a single view or a four-view composite (Doris et al., 2026) to eighty (Su et al., 2015), and view selection ranges from fixed drawing views (Joglekar et al., 2026) through isometric or predetermined angles (M. S. Khan et al., 2024; M. T. Khan et al., 2025) to randomized cameras (Deitke et al., 2023). Datasets rendered under different protocols are difficult to pool or compare view by view, and reconciling them may require re-rendering.

### 4-3. Orientation is rarely reported for CAD data

Only the two foundational studies treat orientation explicitly, by assuming an upright pose or by learning it. Apart from Ortho2CAD, which arranges its views by drafting convention, no CAD-specific source reports how the "front" of a product is determined, and where code is public, models are rendered in their native CAD frame (Doris et al., 2026; Pyatov et al., 2026).

### 4-4. Normalization rules are rarely stated

Among the CAD-specific sources, CADBench reports scaling meshes before rendering (Doris et al., 2026); the released CADFS renderer centers each model and scales its largest extent to one, while the dataset separately supplies physical dimensions for reconstruction (Pyatov et al., 2026); and the released Ortho2CAD generator scales each model so that its largest extent fills a fixed fraction of the view, with one scale shared by its three views, while labelling the true bounding-box dimensions (Joglekar et al., 2026); the other

[Table 1] Requirements for AI-ready multi-view rendering of CAD models and whether designers can meet them without programming

| ID and Section | Requirement | Development-oriented | | | | | | | | Design-oriented | | |
|---|---|---|---|---|---|---|---|---|---|---|---|---|
| | | BlenderProc2 (2023) | Kubric (2022) | pythonOCC (2022) | UnrealCV (2017) | Objaverse-XL script (2023) | Omniverse Replicator (2026) | CAD2Render (2023) | Ortho2CAD generator (2026) | Blender (manual) | KeyShot (manual) | Rhino 8 (manual) |
| Interface needed to operate the tool | | Py | Py | Py | UE+Py | Py | Py | Unity | CLI | GUI | GUI | GUI |
| R1 (4-1, 5-2) | Native STEP, size-relative meshing | - | - | - | M | - | A | - | A | - | A | A |
| R2 (4-4, 5-2) | Uniform scaling to a reference size | - | - | - | - | A | - | M | A | M | M | M |
| R3 (4-4) | Centering, framing, clipping checks | - | - | - | - | - | - | M | A | M | M | M |
| R4 (4-3) | Semantic front and view labels | - | - | - | - | - | - | M | - | M | M | M |
| R5 (4-2) | Fixed view set for all products | - | - | - | - | - | - | - | A | A | A | A |
| R6 (4-2, 5-1) | Shared framing within a product | - | - | - | - | - | - | A | A | A | A | A |
| R7 (4-5) | Orthographic and perspective | - | - | - | - | - | - | - | - | A | A | A |
| R8 (5-1, 5-3) | Clay baseline, matched conditions | - | - | - | - | - | - | - | - | A | M | A |
| R9 (5-3) | Aligned mask, depth, normal passes | - | - | - | - | - | - | A | - | A | A | A |
| R10 (5) | Product-grouped output and splits | - | - | - | - | - | - | - | - | M | M | M |
| R11 (4-4, 5-2) | Per-sample provenance and scale | - | - | - | - | - | - | - | - | M | M | M |
| R12 (4-1, 5-4) | Collection-level (batch) operation | - | - | - | - | - | - | M | A | M | A | M |
| R13 (4-2, 5-4) | Reusable, shareable protocol preset | - | - | - | - | - | - | A | - | A | A | A |
| R14 (4-3, 5-4) | Collection-wide checks and correction | - | - | - | - | - | - | M | - | M | M | M |
| Count | Yes, automatically (A) | 0 | 0 | 0 | 0 | 1 | 1 | 3 | 6 | 6 | 7 | 7 |
| | Yes, manually (M) | 0 | 0 | 0 | 1 | 0 | 0 | 5 | 0 | 7 | 7 | 7 |
| | No (-) | 14 | 14 | 14 | 13 | 13 | 13 | 6 | 8 | 1 | 0 | 0 |

Note. A: automatic for every product once configured. M: manual step repeated for each product / workaround. Dash (-): needs writing or modifying code, or not supported. Ratings reflect the specified nonprogramming user, not overall tool capability. For R12 and R14, M denotes sequential manual workarounds rather than true collection-level functionality.

CAD-specific sources state no rule.

### 4-5. Projection and appearance are chosen per task

Projection is stated in some papers and recoverable only from code in others. Orthographic views appear where the task resembles technical drawing (Joglekar et al., 2026), perspective views where images should resemble photographs (Deitke et al., 2023), and some pipelines mix both (Shi et al., 2026). M. S. Khan et al. (2024) report that perspective distortion degraded their vision-language annotations.

## 5. Requirements

The fourteen requirements are of two kinds. Dataset requirements (R1-R11) define what the images must be: input (R1), geometry (R2-R4), view protocol (R5, R6), observation (R7-R9) and output (R10, R11). They support a model, a human evaluator or both, and their relevance depends on the research question: semantic fronts matter when named views are compared, normalization when relative form is studied, and aligned passes when metrics or validation need them. Interaction requirements (R12-R14) define what the design researcher must be able to do (Section 5-4).

**R4** (semantic front and view labels) is specific to design research: for a designer, "front" is a semantic property of a product rather than a coordinate axis; only if front views show the front of every product can a study ask, for example, whether front views suffice for aesthetic judgment. **R10** (product-grouped output and splits) addresses a general multi-view risk that is easy to overlook: when views of one product are split across training and test sets, a model can score well by recognizing products it has already seen from another angle. Splits for unseen products should also group duplicate models and closely related designs.

### 5-1. Three levels of consistency

Across products, a common scale reference (R2), centering rule (R3) and semantic view convention (R4) make observations comparable. Within each product, shared framing (R6) preserves the relation between views. Across appearance conditions (R8), geometry and cameras are reused while lighting or material changes deliberately.

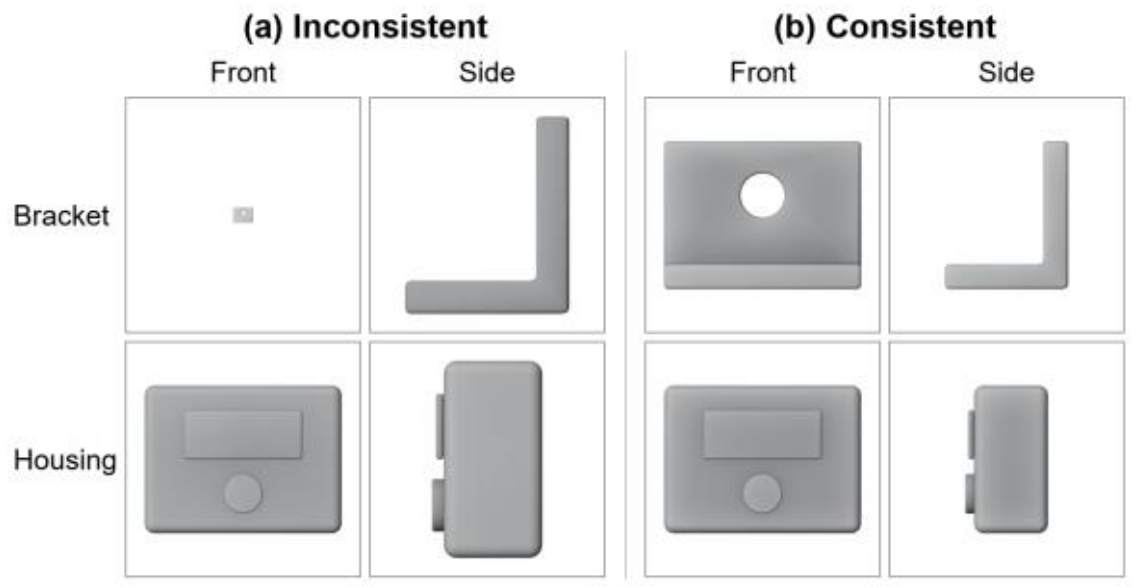

[Figure 1] Consistency across products and within a product

In [Figure 1] (a), a 40 mm bracket and a 400 mm housing are rendered at true size: the front views share one camera, so the bracket almost disappears, while each side view is fitted to its own image, so the housing appears taller from the side than from the front. In [Figure 1] (b), both products are scaled uniformly to the same largest dimension, and all views share one framing.

### 5-2. Proportion rather than absolute size

For studies of relative form, what matters is the size of features relative to the whole rather than their absolute size. A 5 mm fillet reads as soft on a 150 mm remote control but almost sharp on an 800 mm cabinet, and Ho et al. (2016) found that people perceive lower curvature in larger mobile devices. Form indicators are accordingly measured through ratios and distributions rather than absolute dimensions (Ismatullaev & Kim, 2026).

For an image dataset, this argues for normalization that removes absolute size while preserving proportion (R2). Uniform scaling to a declared reference dimension, such as the largest extent or the bounding-box diagonal, preserves every ratio of radius, wall or gap to overall size and reduces disparities in image occupancy. Because absolute size still matters for questions such as ergonomics or scale-aware reconstruction (Pyatov et al., 2026), original dimensions and the scale factor should be stored with each sample (R11). Tessellation tolerance should also scale with product size (R1), so that normalization does not amplify conversion artifacts.

### 5-3. Aligned passes and controlled stimuli

Silhouettes and other form cues can be extracted from RGB images, but aligned passes supply cleaner signals and reference observations (R9). Masks, depth and normals aligned with the RGB image give silhouette, depth and surface direction without segmentation or estimation, which benefits form metrics such as symmetry and curvature character (Ismatullaev & Kim, 2026). For evaluators, rendered images are experimental stimuli (R8): a fixed clay baseline removes material and color cues that would otherwise compete with form, and matched lighting and material variants with stable condition labels let a study such as that of Kim and Kim (2026) separate responses to form from responses to presentation. All passes and conditions of a product remain in the same data split.

### 5-4. Interaction requirements

When the same change must be made product by product, repetition viscosity grows with the collection (Green & Blackwell, 1998), so operations should apply to a whole collection (R12) and the protocol should be stored as one reusable, shareable preset (R13). To support visibility and progressive evaluation, a researcher should be able to inspect every product and view and correct a wrong front, a clipped view or a failed conversion where it is found (R14).

## 6. Comparison of Existing Tools

### 6-1. Three routes, one gap

In the assessed configurations, 31 of the 154 cells are automated, 27 require manual intervention or workarounds, and the rest require programming or are unsupported. The ratings fall into three routes. Ready-made scripts automate a fixed protocol: the Ortho2CAD generator reads a folder of STEP files and normalizes each product to its largest dimension, centers it and frames its three views with one shared scale (R1-R3, R5, R6, R12), and runs without code edits, but changing the protocol requires editing code, so its ceiling is low in the sense of Myers et al. (2000). Frameworks such as BlenderProc2, Kubric, Omniverse Replicator, UnrealCV and pythonOCC reverse the trade-off: they offer building blocks for nearly every requirement, but in the assessed workflows these are configured in code, so their ceiling is high, yet the assumed user can meet almost none.

Design tools offer the third route. In Blender, KeyShot and Rhino, a template scene stores the view set, shared framing, projection and passes as a reusable protocol (R5-R7, R9, R13), but scaling, centering, assigning the front, organizing output, recording provenance and checking results remain manual for each product (R2-R4, R10, R11, R14). KeyShot's bundled batch script renders a folder of CAD files and can center each model on import, but fronts, checks and dataset records remain per-product steps. Four requirements are automatic in no tool: assigning the front (R4), which needs human judgment, product-grouped output with splits (R10), provenance (R11) and collection-wide checks (R14).

### 6-2. Walkthrough of the reference task

The walkthrough identifies two sources of potential interaction cost rather than measuring effort. In design tools, seven requirements retain per-product work even with a template, creating repeated interventions as collections grow or protocols change and increasing opportunities for error.

In scripts and frameworks, a change can be cheap once a script exists, often a small edit and a rerun, but writing and adapting that script is the barrier: in the assessed framework workflows, nearly all requirements are reached only through code, an expertise demand to which the learning barriers of Ko et al. (2004) apply. Threshold and viscosity are therefore different properties, and none of the assessed workflows combines a low threshold with low repetition viscosity while also allowing the protocol to be changed without code.

## 7. Discussion

### 7-1. The gap is scalability, not capability

In Table 1, KeyShot and Rhino provide nonprogramming routes to the reference task, but seven requirements retain per-product work. For R12 and R14, manual ratings represent sequential workarounds rather than true collection-level functionality. No evaluated workflow combines flexible configuration with automated collection-wide preparation and checking without programming. The gap is therefore repeated intervention as collections grow or rendering protocols change.

### 7-2. Implications for a designer-facing tool

The walkthrough points to the need for a designer-facing tool that automates the repetitive steps of the reference task while keeping the flexibility and features that design research needs. Each row of [Table 2] becomes a property of such a tool. The protocol should start from a preset with sensible defaults, while code remains available for extension,

[Table 2] Walkthrough of five actions in the reference task

| Task action | Documented workflow property | HCI concept | Design implication |
|---|---|---|---|
| Set up views, passes and normalization | Frameworks: Python calls. Design tools: a template scene. | Threshold; learning barriers | Preset with defaults; code access kept for extension (R13) |
| Change the normalization rule | Design tools: rescale each product again. Code: a small edit and rerun. | Repetition viscosity | Re-apply the revised preset to all products at once (R12, R13) |
| Assign the front of each product | A per-product judgment, entered as rotations in code or by hand. | Closeness of mapping | Proposed fronts confirmed by direct manipulation in a collection grid (R4, R14) |
| Check framing and clipping | One product at a time in design tools; after the run in scripts. | Visibility; progressive evaluation | Collection-wide preview that flags clipping and failed conversions (R3, R14) |
| Record original size and scale | Design tools: manual rescaling leaves no scale factor with the images. | Hidden dependencies | Write dimensions, scale factor, front and split to every sample by default (R10, R11) |

pairing a low threshold with a high ceiling (Myers et al., 2000). Semantic fronts, which can be ambiguous or task-dependent, should be proposed by the tool and confirmed or corrected by direct manipulation in a view of the whole collection, with an option to mark products that have no meaningful front, as in mixed-initiative interfaces (Horvitz, 1999); the same view should flag clipping and failed conversions. Finally, the tool should record the source, units, transforms, camera parameters, protocol version, original dimensions, scale factor, front and split by default, which supports reporting the rendering protocol alongside dataset documentation (Gebru et al., 2021). These proposals are design hypotheses whose effects on time, errors and workload remain to be evaluated.

### 7-3. Limitations

This paper takes an analytical perspective: the authors carried out the ratings and the walkthrough on one reference task without participants, and the M counts show which requirements retain manual work rather than measured effort. The ten practice sources cover foundational, widely reused and recent CAD-specific work but were not drawn from a systematic search, and the ratings reflect public documentation only.

### 7-4. Future Work

Future work will develop such a designer-facing tool and evaluate both the data, by measuring how controlled variations of normalization, view set, projection and appearance affect learning and human ratings, and the interaction, by comparing manual, scripted and collection-level workflows with design researchers on completion time, errors and workload.

## 8. Conclusion

Rendering CAD models into multi-view images shapes what AI models learn and what evaluators judge, yet it is usually handled by one-off scripts. This paper analyzed reported rendering practice (RQ1), derived fourteen requirements for the dataset and for the interaction that produces it (RQ2), and rated how far design researchers can meet them with eleven tools (RQ3). Existing tools are capable but not scalable: almost every requirement can be met, but through code or through manual steps repeated for every product. These findings motivate collection-level workflows for nonprogramming design researchers.

### Acknowledgement

This work was supported by the Technology Innovation Program (RS-2025-04752997, Development of Design-Production Process Collision Element Prediction and Optimization Solution Technology) funded by the Ministry of Trade, Industry & Energy (MOTIE, Korea)